\documentstyle[11pt,aaspp4]{article}

\def\be{\begin{equation}}
\def\bel#1{\begin{equation}\label{eq:#1}}
\def\ee{\end{equation}}
\def\bea{\begin{eqnarray}}
\def\beal#1{\begin{eqnarray}\label{eq:#1}}
\def\eea{\end{eqnarray}}

\def\eqref#1{\ref{eq:#1}}

\def\gsim{\;\rlap{\lower 2.5pt
 \hbox{$\sim$}}\raise 1.5pt\hbox{$>$}\;}
\def\lsim{\;\rlap{\lower 2.5pt
   \hbox{$\sim$}}\raise 1.5pt\hbox{$<$}\;}
\newcommand\beq{\begin{equation}}
\newcommand\eeq{\end{equation}}
\def\lya{Ly$\alpha$~}

\def\ptilde{{\tilde p}}
\def\Itilde{{\tilde I}}

\def\bfn{{\bf n}}
\def\bfe{{\bf e}}
\def\bfg{{\bf g}}

\begin{document}

\title{Polarization of the Ly$\alpha$ Halos Around Sources Before 
Cosmological Reionization}

\author{George B. Rybicki and Abraham Loeb}
\medskip
\affil{Harvard-Smithsonian Center for Astrophysics, 60 Garden Street,
Cambridge, MA 02138}
%\altaffiltext{2}{email:aloeb@cfa.harvard.edu}

\begin{abstract}

\end{abstract}

In Loeb \& Rybicki (1999; paper I) it was shown that before
reionization, the scattering of \lya photons from a cosmological
source might lead to a fairly compact ($\sim 15^{\prime\prime}$) \lya
halo around the source.  Observations of such halos could constrain
the properties of the neutral intergalactic medium (IGM), and in
particular yield the cosmological density parameters of baryons and
matter on scales where the Hubble flow is unperturbed.  Paper I did
not treat the polarization of this scattered radiation, but did
suggest that the degree of such polarization might be large.  In this
{\it Letter} we report on improved calculations for these \lya halos,
now accounting for the polarization of the radiation field.  The
polarization is linear and is oriented tangentially to the projected
displacement from the center of the source.  The degree of
polarization is found to be 14\% at the core radius, where the
intensity has fallen to half of the central value.  It rises to 32\%
and 45\% at the radii where the intensity has fallen to one-tenth and
one-hundredth of the central intensity, respectively.  At larger radii
the degree of polarization rises further, asymptotically to 60\%.
Such high values of polarization should be easily observable and
provide a clear signature of the phenomenon of \lya halos surrounding
sources prior to reionization.


\section{Introduction}

High--redshift galaxies are detected at present out to $z\la 5.6$,
and are found to be strong \lya emitters (Dey et al. 1998; 
Hu, Cowie, \& McMahon 1998; Spinrad et al. 1998; 
Weymann et al. 1998).
Popular cosmological models predict that at 
somewhat higher redshifts, $z\sim 10$, the hydrogen
in the intergalactic
medium (IGM) was transformed from being predominantly
neutral to being ionized due to the UV radiation
emitted by the first stars and mini-quasars (see, e.g. Gnedin \& Ostriker
1997; Haiman \& Loeb 1998a,b). Prior to this
epoch of  reionization, the
neutral IGM was highly opaque to resonant \lya photons. Hence,
the \lya photons emitted by early galaxies 
were scattered in their vicinity by the surrounding IGM. 
In a previous paper 
(Loeb \& Rybicki 1999, paper I), we have shown that
this intergalactic scattering results 
in compact ($\sim 15^{\prime\prime}$)
halos of \lya light around such sources.
The scattered photons compose a line of a universal
shape, which is broadened and redshifted 
by $\sim 10^3~{\rm km~s^{-1}}$ relative
to the source. The detection of these
intergalactic \lya halos could provide a unique
tool for probing the neutral IGM
before and during the epoch of reionization.
In addition, we have found
that observations of the \lya intensity profile
on scales where the Hubble flow is only weakly
perturbed, 
could constrain the cosmological density parameters of baryons
($\Omega_{\rm b}$) 
and matter ($\Omega_{\rm M}$). 

Paper I had suggested, but not demonstrated,
that the scattered \lya light
would be highly polarized. 
The polarization signal is important in that it 
provides an unambiguous 
signature of the scattering nature
of the diffuse \lya halos around high-redshift
galaxies.
Due to the spherical
symmetry of the scattering geometry, the
linear polarization of the scattered radiation
is expected to be oriented
tangentially relative to the projected displacement from the
center of the source. 

In this {\it Letter} we report on a
detailed calculation of the polarization
properties of 
scattered \lya halos.  In \S 2 we describe 
the Monte-Carlo approach employed in this
calculation, and in \S 3 we describe our 
numerical results. Finally, \S 4 summarizes
the implications of these results.

\section{Polarized Monte Carlo Method}

Let us first discuss the atomic scattering process for the \lya line.
We note that the hydrogen \lya line at $\nu_0=2.466 \times
10^{15}$ Hz is actually a doublet consisting of the two fine-structure lines,
$^2S_{1/2}$ -- $^2P^{\rm O}_{1/2}$ and $^2S_{1/2}$ -- $^2P^{\rm O}_{3/2}$,
separated by $1.1 \times 10^{10}$ Hz.  Fortunately, as shown in paper
I, the regime of interest to us involves frequency shifts from
these line centers of order $\nu_\star \approx 10^{13}$ Hz, which are
much larger than the separation of the lines.  In this regime,
quantum-mechanical interference between the two lines acts in such a
way as to give a scattering behavior identical to that of a classical
oscillator, that is, the same as pure Rayleigh scattering (Stenflo 1980).
(Using an incoherent superposition of the results
of Hamilton [1947] for the two lines, one would incorrectly conclude
that only one-third of the scattering is polarized.)

We shall now describe the modifications of the Monte Carlo method of
paper I necessary to treat polarization.  Monte Carlo methods for
polarized radiative transfer are often formulated using
``photons'' that are actually groups of photons with specified Stokes
parameters (see, e.g., Whitney 1991; Code \& Whitney 1995).  However, for
the present case we found that a more convenient description of
polarization was to use individual photons, each with a definite state
of 100\% linear polarization, a description previously used by Angel
(1969).  (There is no need to consider circular polarization here,
since the central source is assumed to be unpolarized, and Rayleigh
scattering cannot generate circular polarization, except from circular
polarization.)  If the direction of the photon is given by the unit
vector $\bfn$, then its polarization is defined by a real unit vector
$\bfe$, with $\bfn\cdot\bfe=0$.  In this formulation, the observed
Stokes parameters result from the statistics of binning together
multiple, independent photons.

With this description of photons, the polarized Monte Carlo method
involves much the same steps as the unpolarized version of paper I.
One difference is in the handling of the polarized Rayleigh scattering
process, which is done as follows: The angular distribution of the
scattered photon has a probability distribution per solid angle
proportional to $\sin^2 \Theta$, where $\Theta$ is the angle between
the scattered photon and the polarization vector of the incident
photon.  This is simulated using a rejection technique: We choose a
random unit vector $\bfn'$ (uniform in solid angle)
and a uniform random deviate $R$ on
$(0,1)$, and test whether $R < 1-(\bfe \cdot \bfn')^2$; if not, we
start again with new random choices for $\bfn'$ and $R$; the process
is repeated until the test is passed, and then $\bfn'$ is taken as the
new photon direction.  The new polarization vector $\bfe'$ is
determined by finding the normalized projection of the old
polarization vector $\bfe$ onto the plane normal to $\bfn'$, that is,
$\bfe'=\bfg/|\bfg|$, where $\bfg = \bfe -(\bfe \cdot \bfn')\bfn'$.  As
in paper I, an individual photon is followed through a number of
scattering events until it escapes.

The only remaining question is how to characterize the polarization of
the escaped, observed radiation in the plane of the sky as a function
of impact parameter $p$.  From symmetry, we know that this radiation
can be characterized by the intensities parallel to the projected
radius vector, $I_l$, and perpendicular to it, $I_r$.  If $\chi$ is
the angle between the photon's polarization vector and the projected
radius vector, then it contributes to the appropriate histogram bins a
fractional photon number $\cos^2\chi$ to $I_l$ and $\sin^2\chi$ to
$I_r$ (these are the squares of the components of the polarization
vector).  With appropriate normalizations (see paper I), these
histograms determine the two intensities $I_l$ and $I_r$.  In terms of
these the degree of polarization is $\Pi = |I_l -I_r|/(I_l + I_r)$.

An alternative way of stating the results is in terms of the Stokes
parameters $I$ and $Q$, which are related to the above intensities by
$I=I_l+I_r$ and $Q=I_l-I_r$.  The degree of polarization is
$\Pi=|Q|/I$.  These Stokes parameters can also be found directly in
the Monte Carlo method by binning with fractional photon numbers $1$ for
$I$, and $\cos^2\chi-\sin^2\chi=\cos 2\chi$ for $Q$.

The above polarized Monte Carlo method was tested by solving the
classical Milne problem for a Rayleigh scattering atmosphere
(Chandrasekhar 1950; \S68).  The results for the emergent intensities
$I_l$ and $I_r$ agreed, to within statistical errors, with precise
analytical results (Chandrasekhar 1950; table XXIV, p.\ 248).

\section{Results}

As in paper I, we use rescaled variables (denoted by tildes), allowing
one single solution to apply to all physical cases.  In 
particular, we normalize frequencies by,
\begin{equation}
\nu_\star=5.6\times 10^{12}\Omega_{\rm b} h_0
\left[\Omega_{\rm M} (1+z_{\rm s})^{-3} +
(1-\Omega_{\rm M}-\Omega_\Lambda)(1+z_{\rm s})^{-4}
+\Omega_\Lambda(1+z_{\rm s})^{-6}\right]^{-1/2}~
{\rm Hz},
\label{eq:nu_star}
\end{equation}
and distances by,
\begin{equation}
r_\star =
{6.7 (\Omega_{\rm b}/\Omega_{\rm M})~{\rm Mpc} 
\over \left[1+(1-\Omega_{\rm M}-\Omega_\Lambda)
\Omega_{\rm M}^{-1}(1+z_{\rm s})^{-1} + 
(\Omega_\Lambda/\Omega_{\rm M})(1+z_{\rm s})^{-3}\right]},
\label{eq:r_star}
\end{equation}
where $\Omega_{\rm b}$, $\Omega_{\rm M}$
and $\Omega_\Lambda$ are the density 
parameters of baryons, matter, and vacuum, respectively;
$z_{\rm s}$ is the source redshift; and $h_0$ is the Hubble 
constant in units of $100~{\rm km~s^{-1}~Mpc^{-1}}$.
The radiation intensity is normalized by 
$I_\star={\dot N}_\alpha/(r_\star^2\nu_\star)$,
where ${\dot N}_\alpha$ is the steady emission rate
of \lya photons by the source.

The Monte Carlo
method was used to follow the scattering of $10^8$ photons, a
sufficiently large number
to provide reasonable statistical accuracy (except at
very small impact parameters).  The profile of
the frequency-integrated total Stokes intensity $\Itilde$ versus
impact parameter $\ptilde$ is given in Figure 1 (upper solid curve).
The corresponding result from paper I, in which the scattering was
approximated as unpolarized and isotropic, is given as the dashed
curve.  The new curve is seen to be slightly more centrally
concentrated, with a central intensity about 25\% higher than that of
paper I.

The ``core'' radius, where the total Stokes intensity $\Itilde$ has
fallen to half its central value, is at $\ptilde=0.070$.  The radii
where the intensity falls to one-tenth and one-hundredth of the central
intensity are $\ptilde=0.25$ and $0.80$, respectively.  At large
impact parameters, $\Itilde$ falls approximately as $\ptilde^{-3}$.

The polarized intensities $\Itilde_l$ and $\Itilde_r$ are also
plotted in Figure 1.  At the center of the observed disk, $\ptilde=0$,
both of these intensities are equal to $\Itilde/2$ by symmetry.  For all other
values of impact parameter, we note that $\Itilde_r$ exceeds $\Itilde_l$
everywhere, and 
asymptotically by a factor of four.  Thus the radiation is strongly
linearly polarized with orientation tangential to the projected radius
vector.

The lower panel of Figure 1 shows the degree of polarization $\Pi$ versus
impact parameter.  This parameter rises monotonically from zero at
$\ptilde=0$ (as required by symmetry), to about 14\% at the core radius,
and to 32\% and 45\% at the one-tenth and one-hundredth intensity points,
respectively. At still larger impact parameters, the polarization rises
even further to an asymptotic value of 60\%. However, since the intensities
are falling so rapidly, polarizations of that asymptotic magnitude may not
be detectable in practice.

Some heuristic insight into why these polarizations are so high can be
found by comparison to the problem of Rayleigh scattering of radiation from
a point source surrounded by an optically thin scattering medium with
power-law density profile $N(r) \propto r^{-n}$, first treated by Schuster
(1879) in the context of Thomson scattering in the solar corona.  Schuster
showed that the total intensity is a power law in impact parameter, $I
\propto p^{-(n+1)}$, and the degree of polarization is $(n+1)/(n+3)$.  To
relate the asymptotic results of our problem to those of the Schuster
problem, we first note that the radiation at large impact parameters is
dominated by the scattering of photons that have taken one very large step
from much nearer the source, so they are travelling almost radially, as
they would from a point source.  The second observation is that the
Schuster result really applies more generally to a power law dependence of
the product of the scattering cross section $\sigma(r)$ times the density
of the form $\sigma(r)N(r) \propto r^{-n}$.  This is because only the
optical depth along the radial direction is relevant to Schuster's
derivation.  In the situation treated by Schuster, the scattering cross
section was constant.  In our case the density is constant, but the
scattering cross section seen by these radial photons decreases inversely
as the square of the distance, since the frequency displacement is
redshifting linearly with distance, and the line profile varies inversely
as the square of frequency.  Therefore we should compare with the Schuster
results for $n=2$. This gives an inverse third power behavior for the
intensity and $3/5= 60\%$ for the degree of polarization, exactly as seen
in our asymptotic results.

The image of the halo taken in unpolarized light will have circular
symmetry on the sky.  A contour plot of such an image is shown in the
left panel of Figure 2.  The ten contours are spaced by one-half
magnitude ($0.2$ dex), so that the outer contour represents an
intensity $10^{-2}$ times the central intensity.

These contours are distorted when the image is taken through a linear
polarizing filter.  Let us use a cartesian coordinate system
$(\ptilde_x,\ptilde_y)$ in the plane of the sky, and assume the
polarizing filter is oriented along the $\ptilde_y$ axis.  At each
point in the image the observed intensity is given as a weighted average of
the two intensities $\Itilde_l$ and $\Itilde_r$, namely, 
\be
\Itilde(\ptilde_x,\ptilde_y) = \Itilde_l \sin^2 \theta +\Itilde_r \cos^2 \theta,
      \label{eq:iobs} 
\ee 
where $\theta$ is the polar angle of the point relative to the
$\ptilde_x$ axis, so that $\cos \theta = \ptilde_x/\ptilde$ and $\sin
\theta = \ptilde_y/\ptilde$.  Using this equation, the contours of the
halo were constructed and are shown in the right panel of Figure 2,
again using a separation of one-half magnitude between the contours.
The distortions are very evident, even for the innermost contours.

The detectability of \lya halos around high-redshift galaxies was
discussed in detail in Paper I.  We have found that although
difficult, detection of halos at $z_{\rm s}\la 10$ might be feasible
from space.  For example, given the upper limit on the cosmic infrared
background derived by COBE at 1.25 $\mu$m (Hauser et al. 1998), one
could achieve a signal-to-noise ratio ${\rm S/N}=10$ after 10 hours of
integration on an 8-meter space telescope (such as the Next Generation
Space Telescope) for sources at $z_{\rm s}\sim 10$ which possess a Lya
luminosity higher by an order of magnitude than the galaxy discovered
by Dey et al. (1998).

Such sources emit
${\dot{N}}_\alpha= 6\times 10^{54}~{\rm s^{-1}}$, and might be found in
wide-field surveys, based on the broad number-flux distribution which is
predicted for high-redshift galaxies (see Figure 2 in Haiman \& Loeb
1998b).

\section{Conclusions}

As shown in paper I, the special type of \lya halos associated with
sources surrounded by the neutral IGM in a Hubble flow before
reionization, can be clearly characterized by their light profile and
spectrum, which have a universal character.  In this {\it Letter} we
have shown additionally that the polarizations associated with these
\lya halos are quite large, of order tens of percent, and have a
particular universal behavior as a function of impact parameter.

The polarization properties illustrated in Figure 1 could potentially be of
importance in providing a critical test for distinguishing between \lya
halos of the type considered here as opposed to halos due to some other
cause. In addition, the polarization signature can improve the
signal-to-noise ratio in separating faint \lya halos from an unpolarized
background light.

\acknowledgements

This work was supported in part by the NASA grants NAG5-7768 and NAG5-7039
(for AL).  The authors gratefully acknowledge helpful conversations
with Alex Dalgarno and Kenneth Wood.

%-----------------------------FIGURE 1----------------------
\begin{figure}[t]
\centerline{\epsfysize=5.7in\epsffile{ 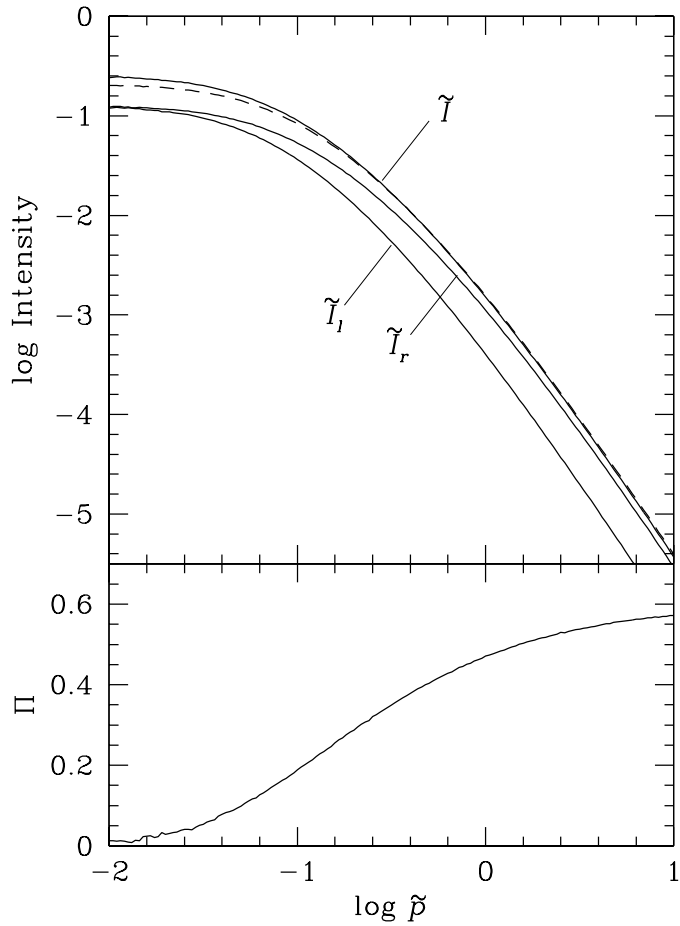 }}
\caption{Upper panel: The solid lines give the frequency integrated
intensities $\Itilde$, $\Itilde_l$, and $\Itilde_r$
versus the impact parameter, $\ptilde$. The dashed
line is the $\Itilde$ found in paper I using isotropic, unpolarized scattering.
Lower panel: The degree of polarization $\Pi$ versus impact
parameter.}
\label{fig:1}
\end{figure}

%-----------------------------FIGURE 2----------------------
\begin{figure}[t]
\centerline{\epsfysize=5.7in\epsffile{ 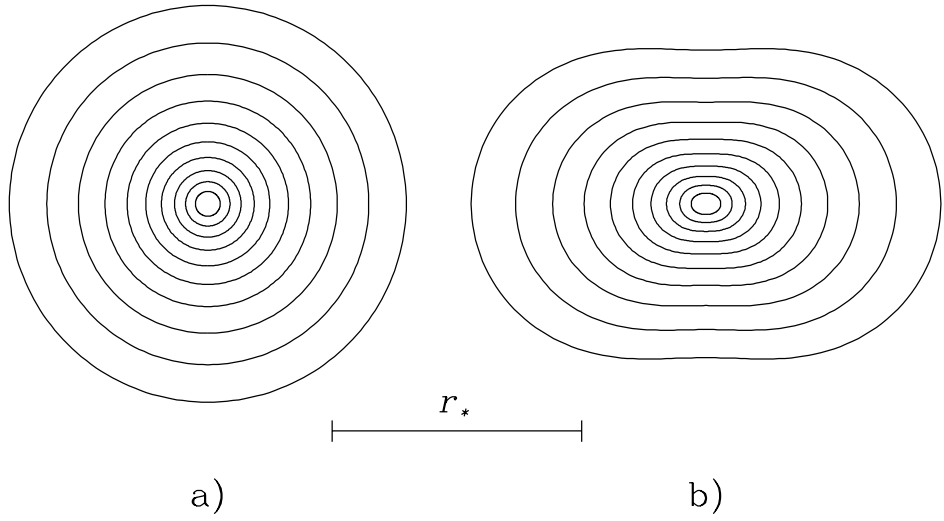 }}
\caption{Predicted images of the \lya halo in the plane of the
sky, as seen in total light (panel a) and
as seen through a linear polarizing filter oriented vertically 
(panel b).
The contours are separated by one-half magnitude, starting with the
central intensity.}
\label{fig:2}
\end{figure}

\end{document}